\documentclass[superscriptaddress,floatfix,prl,aps,twocolumn,10pt]{revtex4-2}
\usepackage{graphicx}
\usepackage{amssymb}
\usepackage{amsfonts}
\usepackage{color}
\usepackage{dsfont}
\usepackage{hyperref}
\usepackage{physics}
\usepackage{xspace}
\usepackage[most]{tcolorbox}
\usepackage[dvipsnames]{xcolor}

\newtcolorbox[]{thmbox}[1][]{breakable,colframe=white,#1}

\usepackage{tikz,xcolor}
\definecolor{lime}{HTML}{A6CE39}
\DeclareRobustCommand{\orcidicon}{%
	\begin{tikzpicture}
	\draw[lime, fill=lime] (0,0) 
	circle [radius=0.16] 
	node[white] {{\fontfamily{qag}\selectfont \tiny ID}};
	\draw[white, fill=white] (-0.0625,0.095) 
	circle [radius=0.007];
	\end{tikzpicture}
	\hspace{-2mm}
}
\foreach \x in {A, ..., Z}{%
	\expandafter\xdef\csname orcid\x\endcsname{\noexpand\href{https://orcid.org/\csname orcidauthor\x\endcsname}{\noexpand\orcidicon}}
}

\begin{document}

\title{Gravitational redshift of broadband relativistic quantum photons}

\author{David Edward Bruschi\orcidB{}}
\affiliation{Institute for Quantum Computing Analytics (PGI-12), Forschungszentrum J\"ulich, 52425 J\"ulich, Germany}
\affiliation{Theoretical Physics, Universit\"at des Saarlandes, 66123 Saarbr\"ucken, Germany}
\email{david.edward.bruschi@posteo.net, d.e.bruschi@fz-juelich.de}
\author{Alessio Lapponi\orcidA{}}
\affiliation{Scuola Superiore Meridionale, Largo San Marcellino 10, 80138 Napoli, Italy}
\affiliation{Istituto Nazionale di Fisica Nucleare, Sezione di Napoli,
Complesso Universitario di Monte S. Angelo, Via Cintia Edificio 6, 80126 Napoli, Italy}
\affiliation{Institute for Quantum Computing Analytics (PGI-12), Forschungszentrum J\"ulich, 52425 J\"ulich, Germany}
\email{alessio.lapponi-ssm@unina.it}
\author{Alessandro Ferreri\orcidC{}}
\affiliation{Center for Quantum Computing, RIKEN, Wakoshi, Saitama, 351-0198, Japan}
\affiliation{Institute for Quantum Computing Analytics (PGI-12), Forschungszentrum J\"ulich, 52425 J\"ulich, Germany}
\email{a.ferreri@fz-juelich.de}
\author{Andreas Wolfgang Schell\orcidD{}}
\affiliation{Division of Light-Matter-Interaction, Johannes Kepler Universität Linz, 4040 Linz, Austria}


\begin{abstract}
Gravitational redshift is a key prediction of general relativity that has been experimentally verified. Nevertheless, its existence in the context of relativistic and quantum physics has not yet been established. 
We employ linearized quantum gravity to study the effects of quantum coherence on gravitational redshift of broadband photons that interact with a planet via graviton exchange. We find that gravitational redshift, as predicted by general relativity, occurs when the initial photonic state contains no quantum coherence between orthogonal photonic modes, such as position states. If quantum coherence is initially present, different photonic modes overlap, a genuine quantum correction occurs solely for single-photon states. Our work not only explains gravitational redshift as the interaction of gravitons and photons in flat spacetime, but also shows that deviations from the theory occur  already at low energies when quantum interference of photonic states is important.
\end{abstract}

\maketitle

\noindent \textbf{INTRODUCTION}

\noindent Gravitational redshift is one of the most well-known predictions of general relativity \cite{Einstein1915,1926Natur.117...86E,FANKHAUSER202419}. An electromagnetic signal travelling in a curved spacetime witnesses a frequency shift between the point of emission and the point of detection. This effect has been experimentally validated \cite{Pound:Rebka:1959,Pound:Rebka:1960,Cranshaw:Schiffer:1960,Will_2014,bothwellResolvingGravitationalRedshift2022,PhysRevLett.45.2081,zhengLabbasedTestGravitational2023, mullerPrecisionMeasurementGravitational2010},  and its impact is now taken into account within many advanced technological applications and tasks, such as satellite clock-synchronization \cite{Wu:Wang:2023}, geodesy \cite{Hackmann:Huckfeldt:2025}, and the global positioning system \cite{Ashby:2002}.

At its core, gravitational redshift is a purely classical prediction that applies to classical light. However, while the electromagnetic field can be found in classical states, such as those used in telecommunications,
photons are intrinsically relativistic \emph{and} quantum particles understood as single exictations of the electromagnetic field that can have quantum states \cite{Srednicki:2007qs}. It is therefore natural to ask if \emph{gravitational redshift occurs as predicted by general relativity also in a relativistic and quantum framework.}

This question is embedded within the ongoing effort of finding signatures of new physics at the intersection of general relativity and quantum mechanics \cite{Karolyhazy:1966,Kiefer:2007,Howl_2019}. Substantial research is currently dedicated to this end, ranging from proposals of a theory of quantum gravity \cite{Rovelli1998,Burgess_2004,Hossenfelder_2013,Marchesano_2024}, stochastic gravity models \cite{Hu:Verdaguer:2008}, and entropy-based approaches \cite{Jacobson:1995,Verlinde:2011,Bianconi:2025}, to semiclassical theories for collapse models \cite{Bassi:Lochan:2013,Bassi:Dorato:2023}, and quantum-classical frameworks \cite{Oppenheim:2023}. Furthermore, many experiments have lately been put forward not only to improve our understanding of physics at the overlap of general relativity and quantum mechanics \cite{Krasnov_2018,Carlesso:Bassi:2019,Roura:2020,Gasbarri:Belenchia:2021,Mohageg:Mazzarella:2022,Mol:Esguerra:2023,Roura:2025,Borregaard:Pikovski:2025}, including using analogue-gravity systems within the laboratory \cite{Barcelo:Liberati:2005}, but also to test the very quantum nature of gravity itself \cite{Ali:Saurya:2011,Marletto:Vedral:2017,Bose:Mazumdar:2017,Carney2021,Lami:Pedernales:2024,Hanif2024,Plavala:2026}. Overall, these avenues employ tools and concepts from quantum information \cite{Nielsen:Chuang:2010,Ralph:Downes:2012}, which has proven to be a successful approach in many related areas of science \cite{Bhattacharyya:Shekar:2018,Camargo:Heller:2019,Kharzeev:2021,Afik:Fabbri:2025}. 

In this work we study gravitational redshift in the context of relativistic and quantum physics. We employ the framework of linearized quantum gravity \cite{Suraj_N_Gupta_1952,Bose_2022}, where the interaction between the electromagnetic and graviton fields occurs in flat spacetime, thus obviating the use of quantum field theory in curved spacetime \cite{Birrell_Davies_1982,Jones_2017,Jones_2018}. This allows us to compare our predictions to those obtained classically by providing a qualitative analysis of the phenomena of interest without loss of generality \cite{Misner:1973prb}. We build on intuition found in a recent quantum-optical model of gravitational redshift affecting photonic quantum states to extend the concept of gravitational redshift to low-energy and weak-gravity relativistic quantum physics \cite{Bruschi:Ralph:2014,Bruschi:Schell:2022,Alan_s_Rodr_guez_2023,molaei2024photongravitycouplingschwarzschildspacetime}. We find that the standard prediction holds for all photonic states except for those initially containing quantum coherence between modes that are orthogonal but overlap. In these cases, a genuine quantum correction appears, thus modifying the classical result.

\vspace{0.4cm}

\noindent \textbf{RESULTS}

\noindent \textbf{Setup modelling}---We study the gravitational redshift of broadband photons in the context of relativistic quantum physics. This requires us to first define gravitational redshift for broadband photons in general relativity.

Let $U_\textrm{A}$ and $U_\textrm{B}$ be the four velocities of two observers Alice and Bob respectively. An ideal pointlike photon with four momentum $P$ that travels between the sender Alice (A) and the receiver Bob (B) has frequency $\omega_K:=-P_\mu U^\mu_K$ as measured locally by each observer $K=\textrm{A},\textrm{B}$. The gravitational frequency shift $z$, hereafter referred to as \emph{gravitational redshift} for simplicity, is given by
\begin{align}\label{gravitationl:redshift:general:relativity}
    1+z=\frac{\omega_\textrm{B}}{\omega_\textrm{A}}\equiv\frac{P_\mu U^\mu_\textrm{B}}{P_\mu U^\mu_\textrm{A}}.
\end{align}
This well-known expression is used whenever the (spectral) size of the photon can be effectively ignored \cite{Misner:1973prb}. However, photons do have a size due to the finiteness of their bandwidth, and this feature is crucial when extending the concept of gravitational redshift to the quantum domain. Thus, we here introduce the \emph{average gravitational redshift} $\bar{z}$ for classical photons with a non-negligible extension:

\begin{thmbox}[boxsep=0pt,left=1em,right=1em,colback=Red!10!white]{}
\textbf{Redshift of extended classical photons}---A broadband classical photon has spatial profile $\varrho(\boldsymbol{x})$.  If the photon witnesses a redshift $z(\boldsymbol{x})$ at each point $\boldsymbol{x}$, we introduce the \textit{average gravitational redshift} 
\begin{equation}\label{average:gravitational:frequency:shift}
    \bar{z}:=\int \textrm{d}^3x\,z(\boldsymbol{x})\,\varrho(\boldsymbol{x})
\end{equation}
witnessed by the photon as a whole.

\vspace{0.1cm}

In this work we focus on setups in weak gravity, where the redshift is small. When the photon is initially localized around $\boldsymbol{x}_0$ far from a static planet, the latter can be approximated as a point-like object and the spacetime can be modelled using the Schwarzschild line element $\textrm{d}s^2=-f(r)\textrm{d}t^2+f^{-1}(r)\textrm{d}r^2+r^2\textrm{d}\Omega^2$, where $f(r)=1-\frac{r_\textrm{S}}{r}$, while $r_\textrm{S}:=2GM_\odot$ is the planet's Schwarzschild radius and $M_\odot$ its mass \cite{Misner:1973prb}. 
In this case, the gravitational redshift $z(\boldsymbol{x}_0)$ for a pointlike photon starting at a distance $|\boldsymbol{x}_0|\gg r_\textrm{S}$ from the planet and traveling far away reads $z(\boldsymbol{x}_0)\approx-\frac{r_\textrm{S}}{2|\boldsymbol{x}_0|}$. Then, the average gravitational redshift $\bar{z}$ for the photon as a whole is 
\begin{equation}\label{average redshift:general:new:main}
    \bar{z}=-\frac{r_\textrm{S}}{2}\int \textrm{d}^3x\,\frac{\varrho(\boldsymbol{x}-\boldsymbol{x}_0)}{|\boldsymbol{x}|}=-\frac{r_\textrm{S}}{2|\boldsymbol{x}_0|}\Gamma_\rho
\end{equation}
where $\Gamma_\rho:=|\boldsymbol{x}_0|\int\textrm{d}^3x\,\frac{\varrho(\boldsymbol{x}-\boldsymbol{x}_0)}{|\boldsymbol{x}|}$ is a scale-independent form-factor that depends only on the spatial shape $\rho$ of the photon. For pointlike photons we have $\varrho(\tilde{\boldsymbol{x}})=\delta^3(\boldsymbol{x})$, and one recovers the textbook expression $z(\boldsymbol{x}_0)=-\frac{r_\textrm{S}}{2|\boldsymbol{x}_0|}$.

\vspace{0.1cm}

If the photon is distributed around multiple locations with a classical profile $\varrho_\textrm{mp}(\boldsymbol{x})=p\varrho(\boldsymbol{x}-\boldsymbol{x}_0)+(1-p)\varrho'(\boldsymbol{x}-\boldsymbol{x}_0')$, then the redshift \eqref{average redshift:general:new:main} becomes
\begin{equation}\label{average redshift:general:new:double:main}
    \bar{z}_\textrm{mp}=-p\,\bar{z}+(1-p)\bar{z}',
\end{equation}
i.e., a weighted average of the classical redshifts.
\end{thmbox}

We now move on to characterize broadband photons as quantum systems. In doing so, we simplify the formalism by considering photons as excitations of scalar fields in order to obtain a qualitative answer to our problem \cite{Misner:1973prb}. 

A broadband and extended photon is a localized field excitation that is created at time $t=0$ by the operator
\begin{align}\label{extended:operator:definition}
    \hat{A}^\dag(F_{\boldsymbol{k}_0}):=\int\textrm{d}^3k F_{\boldsymbol{k}_0}^*(\boldsymbol{k})e^{-i\boldsymbol{k}\cdot\boldsymbol{x}_0}\hat{a}^\dag_{\boldsymbol{k}},
\end{align}
where $\hat{a}_{\boldsymbol{k}},\hat{a}^\dag_{\boldsymbol{k}}$ are the standard annihilation and creation operators that appear in the field decomposition and satisfy the canonical commutation relations $[\hat{a}_{\boldsymbol{k}},\hat{a}^\dag_{\boldsymbol{k}'}]=\delta^3(\boldsymbol{k}-\boldsymbol{k}')$, $F_{\boldsymbol{k}_0}(\boldsymbol{k})$ is the frequency spectrum with characteristic average momentum $\boldsymbol{k}_0:=\int\textrm{d}^3k\, \boldsymbol{k}\,|F_{\boldsymbol{k}_0}(\boldsymbol{k})|^2$ and bandwidth $\sigma$, while  $[\hat{A}(F_{\boldsymbol{k}_0}),\hat{A}^\dag(F_{\boldsymbol{k}_0})]=(F_{\boldsymbol{k}_0},F_{\boldsymbol{k}_0})\equiv\int\textrm{d}^3k |F_{\boldsymbol{k}_0}(\boldsymbol{k})|^2=1$ in order for the photon to have a normalized state.
The one-particle photonic state at time $t=0$ is then given by  $|1_{F_{\boldsymbol{k}_0}}(0,\boldsymbol{x}_0)\rangle:=\hat{A}^\dag(F_{\boldsymbol{k}_0})|0\rangle$.
Within this formalism, we say that a photon initially at $\boldsymbol{x}_0$ is found at time $t$ at position $x(t)$ if $|\langle1_{F_{\boldsymbol{k}_0}}(\boldsymbol{x}(t))|1_{F_{\boldsymbol{k}_0}}(t,\boldsymbol{x}_0)\rangle|^2=1$. It is easy to see using \eqref{extended:operator:definition} that this implies $\boldsymbol{x}(t)=\boldsymbol{x}_0+\frac{\boldsymbol{k}_0}{|\boldsymbol{k}_0|}t$ for localized photons.

The photon interacts gravitationally with a (distant) massive object, i.e., a \emph{planet}. Instead of using the machinery of quantum field theory in curved spacetime \cite{Birrell:Davies:1982}, we employ the well-defined approach of quantized linear gravity \cite{Suraj_N_Gupta_1952}: in such approach the interaction between the field and gravity is obtained via a Lagrangian analysis, where the metric perturbation is an independent quantum field. As a result, the interaction between gravity and any field of interest can be studied using standard quantum field-theoretical tools in flat spacetime \cite{Srednicki:2007qs}. It is crucial to note that gravitons can be present either due to the interaction of matter or photonic fields, or due to an ad-hoc initial state with non-zero average graviton excitations (i.e., a bath of gravitons). While the latter case is often considered in the literature \cite{kannoNoiseDecoherenceInduced2021,boseGravitonsBox2021,torosLossCoherenceCoherence2024}, and can be realized by considering frequencies of the electromagnetic field much higher than those of gravitons \cite{Giovannini:2024}, we here instead assume that gravitons exist only as a consequence of the presence of the planet. 

The (weak) interaction between gravity and the fields of interest is encoded in the interaction Hamiltonian
\begin{equation}
    \hat{H}_{\textrm{I},J}=-\frac{1}{2}\int \textrm{d}^3x\,\hat{h}_{\mu\nu}(\boldsymbol{x}):\hat{T}^{\mu\nu}_J(\boldsymbol{x}):
\end{equation}
where $\hat{h}_{\mu\nu}$ is the quantized metric perturbation field, $:\hat{T}^{\mu\nu}_J(\boldsymbol{x}):$ is the normal ordered stress-energy tensor of the particular field $J$, and $J$ labels the field of interest. 

Gravitational redshift will be obtained in the context of standard Hamiltonian formalism for the time evolution of the reduced photon states. The time-evolution of the whole system is induced by the operator $\hat{U}(t)=e^{-i\hat{H}t}$, where $\hat{H}:=\hat{H}_0+\hat{H}_\textrm{I}$ is the full Hamiltonian, $\hat{H}_0$ is the total free Hamiltonian and $\hat{H}_\textrm{I}=\hat{H}_{\textrm{I},\phi}+\hat{H}_{\textrm{I},\odot}$ is the total interaction Hamiltonian of the gravitational degrees of freedom with the photon and planet respectively.

The next step requires us to specify the initial state $\hat{\rho}(0)$ of the whole system. We assume no initial correlations between the different subsectors (i.e., gravitons, photons, planet) and that no gravitons are initially present, in order to uniquely attribute the effect of interest to the interaction of the subsytems alone. In addition, we model the planet using localized excitations of a massive quantum field with mass $m$ for consistency of treatment. Nevertheless, in the Methods section we show that if we assume that the planet field is initially found in a coherent state $|\Psi_\odot(0)\rangle\equiv|\xi\rangle$ with coherent parameter $\xi\gg1$, the stress-energy tensor of the static planet can be effectively well approximated by the classical expression $T_{\mu\nu}^\odot\approx M_\odot\,\varrho_\odot(\boldsymbol{x})\,\delta_{0\mu}\delta_{0\nu}$, where $M_\odot:=|\xi|^2 m$ is the total mass of the planet, and $\varrho_\odot(\boldsymbol{x})$ is its spatial density distribution. This expression is to be expected for a large classical static planet, and is obtained via a mean-field approach that is common to other areas of physics \cite{Aspelmeyer:Kipppenberg:2014}.

Putting all together, the total initial state of interest reads $\hat{\rho}(0):=|0_\textrm{G}\rangle\langle0_\textrm{G}|\otimes|\Psi_\odot(0)\rangle\langle\Psi_\odot(0)|\otimes\hat{\rho}_\phi(0)$, where $|0_\textrm{G}\rangle$ is the initial reduced state of the gravitons, and $\hat{\rho}_\phi(0)$ is the reduced initial state of the photon. The reduced state $\hat{\rho}_\phi(t):=\textrm{Tr}_{\textrm{G},\odot}(\hat{U}(t)\hat{\rho}(0)\hat{U}^\dag(t))$ of the photon at time $t$ can be computed and we find
\begin{align}\label{reduced:state:time:evolution}
    \hat{\rho}_\phi(t)\approx \hat{U}_\phi(t)\hat{\rho}_\phi(0)\hat{U}^\dag_\phi(t),
\end{align}
where the effective time-evolution operator $\hat{U}_\phi(t):=\overset{\leftarrow}{\mathcal{T}}\exp\bigl[-i \int\textrm{d}t'\bigl(\hat{H}_{0,\phi}+\hat{H}_{2,\phi}(t')\bigr)\bigr]$ is induced by the \emph{effective classical-quantum gravitational Hamiltonian}
\begin{align}
\hat{H}_{2,\phi}(t)=&r_\textrm{S}\int \frac{\textrm{d}^3k\textrm{d}^3k'}{2\pi^2(2\pi)^3}\mathcal{H}_{\boldsymbol{k},\boldsymbol{k}'}(t)\tilde{\rho}_\odot(\boldsymbol{k}-\boldsymbol{k}')\hat{a}^\dagger_{\boldsymbol{k}}\hat{a}_{\boldsymbol{k}'},
\end{align}
which acts on the reduced space of the photons. Here we have introduced the function $\mathcal{H}_{\boldsymbol{k},\boldsymbol{k}'}(t)=-\frac{\sqrt{|\boldsymbol{k}||\boldsymbol{k}'|}}{|\boldsymbol{k}-\boldsymbol{k}'|^2}\bigl(1-\cos\bigl(|\boldsymbol{k}-\boldsymbol{k}'|t\bigr)\bigr)$ for convenience, the Fourier transform $\tilde{\rho}_\odot(\boldsymbol{k})=\int \textrm{d}^3x\varrho_\odot(\boldsymbol{x})e^{i\boldsymbol{x}\cdot\boldsymbol{k}}$ with $\tilde{\rho}_\odot(\boldsymbol{k})=\tilde{\rho}_\odot^*(\boldsymbol{k})$,  and we have employed the rotating wave approximation to cancel co-rotating terms \cite{Heib:Lageyre:2025}. Note that $\hat{H}_{2,\phi}(t)$ is a collection of beam-splitters and phase rotations \cite{Scully:1997}, i.e., \emph{passive operations} \cite{Adesso:Ragy:2014}. All details are left to the Methods.

\begin{figure}[t]
    \includegraphics[width=0.9\linewidth]{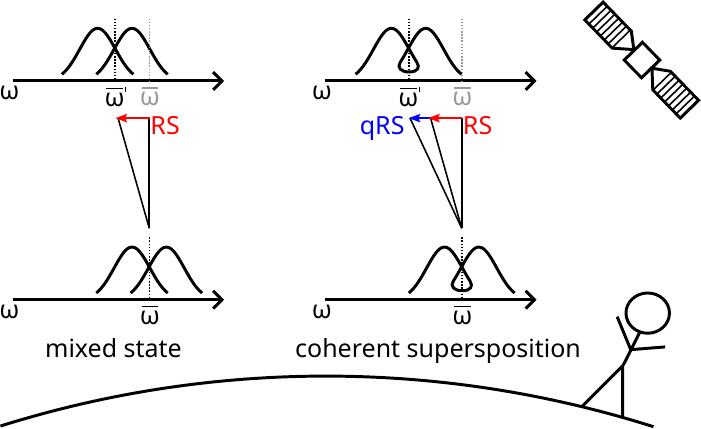}
    \caption{\textbf{Schematic figure of the setup}. Photons propagating in weakly curved spacetime are viewed as field excitations that interact with a classical massive source (e.g., planet) via graviton exchange. RS denotes the redshift and qRS the quantum redshift while $\bar{\omega}$ is the average frequency and $\bar{\omega}'$ is the redshifted frequency.}
    \label{fig:one}
\end{figure}

\vspace{0.8cm}

\noindent \textbf{Defining a relativistic quantum redshift}---We have set up the machinery to obtain our goal, but we are now faced with a fundamental question:

\vspace{0.1cm}

\noindent\textbf{Q}: \emph{Under which conditions is redshift present in the quantum state of a broadband photon interacting with gravity}?

\vspace{0.1cm}

\noindent This question is equivalent to asking how can we define gravitational redshift in the context of relativistic and quantum physics. Clearly, any proposed definition must capture the standard expression \eqref{average:gravitational:frequency:shift} in the limit of classical localized photons. Since we cannot provide a general definition that is independent of the boundary conditions of the problem, we choose an operational approach that is motivated by concrete practical implementations and that reproduces the expected results in the classical limit. 

The intuition behind our proposal stems form the results obtained in recent work on this topic \cite{Bruschi:Ralph:2014,Bruschi:Schell:2022,Alan_s_Rodr_guez_2023}: namely, in the case of weak gravity, the (average) energy $\bar{E}$ of a photon is well approximated by the relation $\bar{E}=\hbar\,\bar{\omega}$, where $\bar{\omega}:=\frac{c}{\hbar}\int \textrm{d}^3k\,|\boldsymbol{k}||F_{\boldsymbol{k}_0}(\boldsymbol{k})|^2$ is its \emph{average} frequency. It has been shown that $\bar{\omega}$ is expected to transform according to Eq.~\ref{gravitationl:redshift:general:relativity} at least in the case of highly confined photons along the direction of propagation, where the system can effectively be treated as one dimensional and therefore the frequency profile $F(\omega)$ transforms as $F_\textrm{A}(\omega)\rightarrow F_\textrm{B}(\omega)=(1+z)^{-1/2}F_\textrm{A}(\omega/(1+z))$ from the local point of view of the receiver \cite{Alan_s_Rodr_guez_2023}. Thus, the energy $\bar{E}=c\int \textrm{d}^3k\,|\boldsymbol{k}||F_{\boldsymbol{k}_0}(\boldsymbol{k})|^2$ must also transform as a result of gravitational redshift exactly in the same fashion that the frequencies do, cf. Eq.~\ref{gravitationl:redshift:general:relativity}, due to the relations $\bar{E}=\hbar\,\bar{\omega}$ and $\bar{E}_\textrm{B}/\bar{E}_\textrm{A}=\bar{\omega}_\textrm{B}/\bar{\omega}_\textrm{A}$. We take this latter transformation to provide our classical benchmark for the relativistic and quantum extension. 
Finally, since we are working in the context of weak gravity, the redshift is expected to have the generic expression \eqref{average redshift:general:new:main}. A pictorial representation of the scheme is given in Figure~\ref{fig:one}.

We now return to the Hamiltonian formalism. In the reduced space of the photon with reduced state $|\Psi(t)\rangle$, we introduce the average instantaneous energy 
\begin{equation}\label{average:energy}
E(t):=\langle\Psi(t)|i\,\hbar\,\partial_t|\Psi(t)\rangle\equiv\langle\Psi(t)|\hat{H}_\phi(t)|\Psi(t)\rangle,
\end{equation} 
which reads $E(t)=\textrm{Tr}(\hat{H}_{\phi}(t)\hat{\rho}(t))$ for mixed states.
Given the discussion above, we define:

\begin{thmbox}[boxsep=0pt,left=1em,right=1em,colback=NavyBlue!10!white]{}
\textbf{Redshift of broadband quantum photons}---Let $\bar{z}_\textrm{AB}=-\frac{r_\textrm{S}}{2|\boldsymbol{x}_\textrm{A}|}\Gamma_{\rho,\textrm{A}}+\frac{r_\textrm{S}}{2|\boldsymbol{x}_\textrm{B}|}\Gamma_{\rho,\textrm{B}}$ be the classical redshift associated to a classical photon with form-factor $\Gamma_{\rho,K}:=|\boldsymbol{x}_K|\int\textrm{d}^3x\,\frac{\varrho(\boldsymbol{x}-\boldsymbol{x}_K)}{|\boldsymbol{x}|}$ that is sent at time $t_\textrm{A}$ by Alice, located at $\boldsymbol{x}_\textrm{A}$, and is received at time $t_\textrm{B}$ by Bob, located at $\boldsymbol{x}_\textrm{B}$. Then, the photon has witnessed an \emph{effective redshift} $\bar{z}_\textrm{AB}$ if
\begin{equation}\label{relativistic:quantum:redshift}
    \frac{E(t_\textrm{B})}{E(t_\textrm{A})}=1+\bar{z}_\textrm{AB}.
\end{equation}
\end{thmbox}
In other words, a photon that is initially prepared in an arbitrary quantum state $\hat{\rho}(0)$ is redshifted if and only if its change of average energy matches the variation expected classically due to redshift. This guarantees that, at least in the case of classical photons, we should recover the expected effects predicted by general relativity.

\vspace{0.1cm}

\noindent \textbf{Redshift of broadband quantum photons}---We now have all the tools to compute the effects of interest. Therefore, we can proceed as follows: we start by computing the effect in the case of a single photon in a single-mode state that can be completely mixed or pure. We then allow for an arbitrary one-photon single-mode state that can contain quantum coherence. This approach allows us to compare the results in a meaningful way. The boundary conditions of the problem will be the following: a photon is emitted from $\boldsymbol{x}_0\gg r_\textrm{S}$ at $t=0$ and subsequently travels to a final location $\boldsymbol{x}(t)=\boldsymbol{x}_0+\frac{\boldsymbol{k}_0}{|\boldsymbol{k}_0|} t$ where it is detected at time $t$. The modes that define the initial photonic states are given by the mode functions $F_{\boldsymbol{k}_0}$ and $F_{\boldsymbol{k}_0}'$ with parameters $(\boldsymbol{k}_0,\sigma,\boldsymbol{x}_0)$ and $(\boldsymbol{k}_0',\sigma',\boldsymbol{x}_0')$, which determine the two orthogonal single-particle states $|\Psi(0)\rangle:=\hat{A}^\dag(F_{\boldsymbol{k}_0})|0\rangle$ and $|\Psi'(0)\rangle:=\hat{A}^\dag(F_{\boldsymbol{k}_0}')|0\rangle$ that satisfy $\langle \Psi(0)|\Psi'(0)\rangle=\int\textrm{d}^3k\,F_{\boldsymbol{k}_0}^*(\boldsymbol{k})F_{\boldsymbol{k}_0}'(\boldsymbol{k})=0$.
Finally, we assume that the photon is very peaked as compared to its average frequency (or momentum), i.e.,  $\frac{\sigma}{|\boldsymbol{k}_0|},\frac{\sigma'}{|\boldsymbol{k}_0'|}\ll1$. The consequences of this assumption are discussed below.

\vspace{0.1cm}

\noindent \textit{Single-photon single-mode mixed state}---We start with a single photon initially in a mixed state. It is well known that mixed states in the sharp momentum basis in quantum field theory are ill-defined \cite{Bruschi:Chatzinotas:2021}. Thus, we employ a method developed in the literature to circumvent this problem. Let $F_{\boldsymbol{k}_0}(\boldsymbol{k})=f_{\boldsymbol{k}_0}(\boldsymbol{k})e^{i\psi(\boldsymbol{k})}$, where $f_{\boldsymbol{k}_0}(\boldsymbol{k})=|F_{\boldsymbol{k}_0}(\boldsymbol{k})|$. We then introduce the state
\begin{equation}\label{single:mode:mixed:state}
    \hat{\rho}_\textrm{m}(0)=\sum_n f_{\boldsymbol{k}_0}^2(\boldsymbol{k}_n)|1_{\boldsymbol{k}_n}\rangle\langle1_{\boldsymbol{k}_n}|,
\end{equation}
where $|1_{\boldsymbol{k}_n}\rangle$ are rectangle wavepackets that form an approximate orthonormal basis and are very small as compared to the photonic wavepacket, i.e., they have width $\lambda\ll\sigma$. Using these states we can write $|\Psi(0)\rangle=\sum_n F_{\boldsymbol{k}_0}(\boldsymbol{k}_n)|1_{\boldsymbol{k}_n}\rangle+\mathcal{O}(\lambda/\sigma)$, an approximation that improves with decreasing $\lambda$. Thus, the state $\hat{\rho}_\textrm{m}(0)$ has been effectively obtained from the state $|\Psi(0)\rangle\langle\Psi(0)|$ by removing the off-diagonal terms \cite{Bruschi:Chatzinotas:2021}.

We use the fact that $E(0)=|\boldsymbol{k}_0|$, compute the average energy \eqref{average:energy}, and find
\begin{equation}\label{gravitational:redshift:mixed:single:mode}
    \frac{E(t)}{E(0)}=1-\frac{r_\textrm{S}}{2|\boldsymbol{x}_0|}\Gamma_\rho+\frac{r_\textrm{S}}{2|\boldsymbol{x}(t)|}\Gamma_\rho(t),
\end{equation}
where $\boldsymbol{x}(t)=\boldsymbol{x}_0+\frac{\boldsymbol{k}_0}{|\boldsymbol{k}_0|}t$ and $\Gamma_{\rho}(t):=|\boldsymbol{x}(t)|\int\textrm{d}^3x\,\frac{\varrho(\boldsymbol{x}-\boldsymbol{x}(t))}{|\boldsymbol{x}|}$ are the position and form factor of the photon at time $t$, respectively. The expression \eqref{gravitational:redshift:mixed:single:mode} has the required form \eqref{relativistic:quantum:redshift}, where $\bar{z}_\textrm{cl}(t)=-\frac{r_\textrm{S}}{2|\boldsymbol{x}_0|}\Gamma_\rho+\frac{r_\textrm{S}}{2|\boldsymbol{x}(t)|}\Gamma_\rho(t)$ coincides with the classical redshift expression \eqref{average redshift:general:new:main} for the case at hand. 

This first result is encouraging because it reinforces the definition \eqref{relativistic:quantum:redshift} of redshift for wave packets of relativistic and quantum fields, since it reduces to the expected classical expression in the case of mixed, classical states. 

\vspace{0.1cm}

\noindent \textit{Single-photon single-mode pure state}---We proceed with considering the initial pure state $|\Psi(0)\rangle$ for the photon. We then compute the quantity $\frac{E(t)}{E(0)}-1$ for this case and find that it coincides with that of the previous case, i.e., with expression \eqref{gravitational:redshift:mixed:single:mode}.
This tells us that single photons with a single-peaked and narrow-band frequency profile $F_{\boldsymbol{k}_0}$ witness redshift as predicted in general relativity.

\vspace{0.1cm}

\noindent \textit{Single-photon two-mode state}---We finally consider the most generic single-photon two-mode state defined by two orthogonal modes $|\Psi(0)\rangle$ and $|\Psi'(0)\rangle$. It reads
\begin{align}\label{initial:state}
    \hat{\rho}(p,q)=&p|\Psi(0)\rangle\langle\Psi(0)|+(1-p)|\Psi'(0)\rangle\langle\Psi'(0)|\nonumber\\
    +&q\,e^{-i\theta}|\Psi(0)\rangle\langle\Psi'(0)|+q \,e^{i\theta}|\Psi'(0)\rangle\langle\Psi(0)|.
\end{align}
This state has been recently used in studies of the role of quantum coherence in relativistic and quantum processes \cite{Bruschi:2016,Lapponi:Mancini:2026}. In order to have a valid physical state we must require $0\leq p\leq1$, that $q$ is real and bounded by $-1/2\leq q\leq1/2$, and finally $(p-1/2)^2+q^2\leq1/4$. With these constraints, we note that the state is maximally mixed for $p=1/2$ and $q=0$, while it is maximally entangled for $q=p=1/2$. The phase $\theta$ is unconstrained.

The parameters $p$ and $q$ have important physical meaning. The former is the probability $p$ of detecting the photon in mode $|\Psi(0)\rangle$ (and therefore probability $1-p$ of detecting the photon in the orthogonal mode $|\Psi'(0)\rangle$). The latter parameter is related to the amount of quantum coherence in the state. In general, an arbitrary quantum state $\hat{\rho}=\sum_{jk} \rho_{jk}\ket{j}\bra{k}$, where the states $\ket{j}$ form an orthonormal basis, contains quantum coherence. This can be seen using the \emph{ relative entropy of coherence} $C (\hat{\rho}):=S(\hat{\rho}_\textrm{diag})-S(\hat{\rho})$, where $S(\hat{\rho}):=-\textrm{Tr}(\hat{\rho}\ln\hat{\rho})$ is the von Neumann entropy of state $\hat{\rho}$ and $\hat{\rho}_\textrm{diag}$ is the state obtained from $\hat{\rho}$ by removing all off-diagonal terms \cite{Zhang:Shao:2016}. In our case we have $C(p,q)\equiv C(\hat{\rho}(p,q)) =-\sum_{\sigma=\pm}\left[\lambda^\textrm{diag}_\sigma\ln\lambda^\textrm{diag}_\sigma-\lambda_\sigma\ln\lambda_\sigma\right]$,
which is supplemented by the eigenvalues $\lambda^\textrm{diag}_+=p$, $\lambda^\textrm{diag}_-=(1-p)$, and $\lambda_\pm=\frac{1}{2}\bigl[1\pm\sqrt{(1-2\,p)^2+4\,q^2}\bigr]$.
It is easy to verify that $C(p,q)$ is a monotonic function of $q$, with minimal value $C(p,0)=0$ for arbitrary $p$ and maximal value $C(1/2,1/2)=\ln2$. 

We now compute the average energy $E_{pq}(t)=\text{Tr}(\hat{H}(t)\hat{U}(t)\hat{\rho}(p,q)\hat{U}^\dag(t))$ as well as the initial energy $E_{p,q}(0)=|\boldsymbol{k}_0|$ for our current case, and find
\begin{align}\label{two:mode:state:average:energy}
    \frac{E_{p,q}(t)}{E(0)}=&1+p\,\bar{z}(t)+(1-p)\,\bar{z}'(t)+q\,\bar{w}(t),
\end{align}
where $\bar{z}(t)$ and $\bar{z}'(t)$ have the standard classical redshift expressions \eqref{gravitational:redshift:mixed:single:mode} with $\boldsymbol{x}_0$ and $\boldsymbol{x}_0'$ replaced by $\boldsymbol{x}(t):=\boldsymbol{x}_0+\frac{\boldsymbol{k}_0}{|\boldsymbol{x}_0|}t$, and $\boldsymbol{x}'(t):=\boldsymbol{x}'_0+\frac{\boldsymbol{k}_0}{|\boldsymbol{x}_0|}t$ respectively, while the additional new cross-term $\bar{w}(t)$ has the expression
\begin{align}\label{crucial:cross:term}
    \bar{w}(t)=&-\frac{r_\textrm{S}}{2}\textrm{Re}\left\{e^{i(\psi_0-\theta)}\left[\int\textrm{d}^3x\frac{\tilde{F}(\boldsymbol{x}-\boldsymbol{x}_0)\tilde{F}'{}^*(\boldsymbol{x}-\boldsymbol{x}_0')}{|\boldsymbol{x}|}\right.\right.\nonumber\\
    &-\left.\left.\int\textrm{d}^3x\frac{\tilde{F}(\boldsymbol{x}-\boldsymbol{x}(t))\tilde{F}'{}^*(\boldsymbol{x}-\boldsymbol{x}'(t))}{|\boldsymbol{x}|}\right]\right\}.
\end{align}
Here we have assumed that the photon departs and is aligned outward radially, i.e., $\boldsymbol{k}_0,\boldsymbol{x}_0,\boldsymbol{x}_0'$ are aligned, to simplify the expressions. The phase $\psi_0:=\boldsymbol{k}_0\cdot(\boldsymbol{x}_0-\boldsymbol{x}_0')$ is not important and will be discarded, since it can be absorbed by a redefinition of the mode functions. 

We now comment this last, and most important, result. First of all, it is immediate to see that the ``redshift'' for the state $\hat{\rho}(p,q)$ of interest is 
\begin{equation}
    \bar{z}_{p,q}(t)=p\,\bar{z}(t)+(1-p)\,\bar{z}'(t)+q\,\bar{w}(t).
\end{equation}
When $p=1$ and $q=0$ we recover the classical redshift $\bar{z}_{1,0}(t)=\bar{z}_\textrm{cl}(t)$ expected for the photon starting in the state $|\Psi(0)\rangle$, while when $p=0$ and $q=0$ we recover the classical redshift $\bar{z}_{0,0}(t)=\bar{z}'(t)$ expected for the photon starting in the state $|\Psi'(0)\rangle$. Thus, when no quantum coherence is present, standard general relativity applies. 

We then look at the more interesting case $q>0$, where quantum coherence is initially present in the state. Note that, in such case, $0<p<1$ due to the constraints presented above for the state to be viable, and thus the photon can be found with probability $p$ or $1-p$ in mode $|\Psi(0)\rangle$ or $|\Psi'(0)\rangle$ respectively. The term \eqref{crucial:cross:term} contributes in this case, and it is uniquely determined by the overlap expression $R_{\boldsymbol{x},\boldsymbol{x}'}(t)=\tilde{F}(\boldsymbol{x}-\boldsymbol{x}(t))\tilde{F}'{}^*(\boldsymbol{x}-\boldsymbol{x}'(t))$ as a function of time $t$, for all times.

We now note the following. If the functions $\tilde{F}$ and $\tilde{F}'$ have no common support, for example are Gaussians localized far away from each other, then $\bar{w}(t)=0$ and the overall redshift $\bar{z}_{p,q}(t)$ at time $t$ coincides with what one would expect classically: a linear combination $\bar{z}_{p,q}(t)=p\,\bar{z}(t)+(1-p)\,\bar{z}'(t)$ of the redshifts $\bar{z}(t)$ and $\bar{z}'(t)$ of the individual modes weighted accordingly. However, when the functions $\tilde{F}$ and $\tilde{F}'$ do have common support, then $\bar{w}(t)$ can be nonzero, and thus the overall redshift $\bar{z}_{p,q}(t)$ at time $t$ deviates from what is expected classically. This particular instance can occur, for example, when the two mode functions $\tilde{F}$ and $\tilde{F}'$ are given a complex phase that enables their orthogonality. 

The conclusions above would apply \emph{mutatis mutandis} even if the photon modes were characterized by different initial average momenta $\boldsymbol{k}_0$ and $\boldsymbol{k}_0'$. In such case, one would have $E_{pq}(t)=E(0)+p\,\bar{z}(t)|\boldsymbol{k}_0|+(1-p)\,\bar{z}'(t)|\boldsymbol{k}_0'|+q\,\bar{w}(t)\sqrt{|\boldsymbol{k}_0||\boldsymbol{k}_0'|}$, with $E_{pq}(0)=p|\boldsymbol{k}_0|+(1-p)|\boldsymbol{k}_0'|$.

\vspace{0.1cm}

\noindent \textit{Quantifying the effects}---We now turn to the magnitude of the effects predicted in this work. We focus on the simplest possible scenario, that of a photon traveling far away from the planet, which implies $|\boldsymbol{x}(t)|,|\boldsymbol{x}'(t)|\rightarrow+\infty$. In such case we are left with the constant terms
{\small
\begin{subequations}\label{simple:case}
    \begin{align}
    \bar{z}(t)=-\frac{r_\textrm{S}}{2|\boldsymbol{x}_0|}\Gamma_\rho,\qquad \bar{z}'(t)=-\frac{r_\textrm{S}}{2|\boldsymbol{x}_0'|}\Gamma_\rho',
    \end{align}
    \begin{align}
    \bar{w}(t)=-\frac{r_\textrm{S}}{2}\textrm{Re}\left[e^{-i\theta}\int\textrm{d}^3x\frac{\tilde{F}(\boldsymbol{x}-\boldsymbol{x}_0)\tilde{F}'{}^*(\boldsymbol{x}-\boldsymbol{x}_0')}{|\boldsymbol{x}|}\right].
\end{align}
\end{subequations}
}

\noindent Since the mode functions are square integrable and peaked initially around $\boldsymbol{x}_0$ and $\boldsymbol{x}_0'$, it is clear that $\bar{w}(t)$ can be non-negligible only when the distance $\boldsymbol{L}:=\boldsymbol{x}_0'-\boldsymbol{x}_0$ is not much larger than the characteristic size $\ell_0\sim1/\sigma$ of the photon as argued before. Otherwise, when $\ell_0/L\ll1$ with $L:=|\boldsymbol{L}|$, the gravitational redshift coincides with the classical value $\bar{z}_{p,q}(t)=p\,\bar{z}(t)+(1-p)\,\bar{z}'(t)$. Operationally, this means that an observer would measure the redshift $\bar{z}(t)$ a fraction $p$ amount of times, while he would measure $\bar{z}'(t)$ a fraction $(1-p)$ amount of times over a statistically large-enough sample, if no other knowledge of the photon was available.

The situation is different when $L\approx\mathcal{O}(\ell_0)$. In such case, the cross-term $\bar{w}(t)$ can become non-negligible and the redshift can deviate from the classical value. Crucially, this contribution is purely quantum in nature since, as we have argued above, $q\neq0$ only when quantum coherence is present in the initial state. The triangle and Cauchy-Schwarz inequalities give us the ultimate upper bound $|\bar{w}(t)|^2\leq\bar{z}(t)\cdot\bar{z}'(t)$ on the magnitude of the quantum contribution.

We compute the quantities \eqref{simple:case} for the case where $\boldsymbol{x}_0'=\boldsymbol{x}_0$, and find $\bar{z}'(t)=\bar{z}(t)$ as well as
\begin{equation}
    \bar{w}(t)=\frac{1}{2}\frac{r_\textrm{S}}{|\boldsymbol{x}_0|}\frac{\ell_0}{|\boldsymbol{x}_0|}\Gamma_w(\theta),
\end{equation}
where we have defined the form factor
\begin{equation}
    \Gamma_w(\theta):=\textrm{Re}\left[e^{-i\theta}\int\textrm{d}^3w\,\boldsymbol{w}\cdot\frac{\boldsymbol{x}_0}{|\boldsymbol{x}_0|}\tilde{F}(\boldsymbol{w})\tilde{F}'{}^*\bigl(\boldsymbol{w}\bigr)\right]
\end{equation}

\noindent Here, $\boldsymbol{w}$ is a dimensionless variable, and the modes $F,F'$ are orthogonal due to the presence of a complex phase.

It is clear that, for broadband  photons whose length $\ell_0$ is much smaller than the distance $|\boldsymbol{x}_0|$ at which the photon is initially located, the quantum correction $\bar{w}(t)$ is of order $\mathcal{O}(\ell_0/|\boldsymbol{x}_0|)$ smaller than the classical redshift itself. For example, for a $6$~m optical photon created at the Earth's surface we have $\ell_0/|\boldsymbol{x}_0|\approx10^{-6}$. Thus, measuring the effects in Earth-to-satellite experiments remains challenging with current technologies. New methods must be developed to enhance the effect, rendering it detectable.

It might seem desirable to use states with multiple photons. Nevertheless, it is straightforward to verify that, if we were to use a state of the form \eqref{initial:state} with an $n_{F}$ particle state $|\Psi(0)\rangle=|n_{F}\rangle$ of mode $F_{\boldsymbol{k}_0}$ and an $m_{F'}$ particle state $|\Psi'(0)\rangle=|m_{F'}\rangle$ of mode $F_{\boldsymbol{k}_0}'$, we would then have $\bar{w}(t)\equiv0$. This induces us to conclude that, surprisingly, quantum corrections to small redshift can appear only for single-photon states, or states defined by superpositions that include single-photon ones, such as coherent states \cite{Adesso:Ragy:2014}.
We summarize the outcome of our work in Table~\ref{tab:my_label}.

It is worth noting the role of the phase parameter $\theta$ in \eqref{crucial:cross:term}: its main non-trivial effect is to change the sign of $\bar{w}(t)$, i.e., decrease or enhance the effective classical redshift, which is unexpected. This is directly related to recent results showing that the phase of a quantum system in a superposition of different locations can effectively result in a small repulsive gravitational force \cite{Saldanha:martletto:2026}. More work is necessary to explore this connection.

\begin{table}[ht!]
    \begin{tabular}{|c||c|c|}
        \hline
           $z(t)$        & \textbf{Single-Mode} & \textbf{Two-Mode}\\
       \hline
       \hline
       \textbf{Classical}  & $\bar{z}_\textrm{cl}(t)$ & $\bar{z}_{2,\textrm{cl}}(t)=p\bar{z}_\textrm{cl}(t)+(1-p)\bar{z}_\textrm{cl}'(t)$ \\
       \hline
       \textbf{Quantum}  & $\bar{z}_\textrm{cl}(t)$ & $\bar{z}_\textrm{quant}(t)=\bar{z}_{p,q}(t)$\\
       \hline
    \end{tabular}
    \caption{\textbf{Redshift for single broadband photons}: the generic form of the redshift is presented as a function of the number of modes in the state, and of the quantum coherence contained in the state. The classical result can be expected in all cases except in the case of a  superposition of two orthogonal single-mode states that have spatial overlap.}
    \label{tab:my_label}
\end{table}
We conclude by noting that the results presented here rely on the photon being narrow-band, since corrections to all quantities occur at order $\mathcal{O}(\sigma/|\boldsymbol{k}_0|)$. If the photons have a spectral density that is of comparable size to the peak frequency, then the classical redshift cannot be recovered. A detailed numerical analysis is then required on a case-by-case basis since no general analytical expression can be obtained in such case.

\vspace{0.4cm}

\noindent \textbf{DISCUSSION}

\noindent We have studied gravitational redshift of broadband photons in the context of low-energy and weak-curvature relativistic and quantum physics. This required us to first introduce redshift for classical broadband photons within general relativity, and then to extend the definition to the relativistic and quantum case in such a way that the classical result is obtained in the appropriate limit. We achieved our goal by shifting the problem of determining the dynamics of quantum fields in a classical weakly curved spacetime to that of interacting fields, including the graviton field, in flat spacetime. 

We find that the classical redshift is recovered for narrow-band photons in the case where no quantum coherence is contained in the initial state, while a correction that is purely quantum in nature occurs for states with quantum coherence between different photonic modes with nonvanishing overlap in the spatial (or equivalently frequency) domain. Furthermore, we find that the quantum correction is a genuine single photon effect, while genuine multi-photon states can witness only a classical redshift. Understanding this peculiar single-photon behavior requires further investigation.

The effects predicted here scale as the small photon-length-to-distance ratio. High sensitivity is therefore necessary do detect such a correction in near-Earth experiments on top of the already
small local gravitational redshift. While current technologies and advanced experimental systems have shown great sensitivity to small phase changes within interferometers \cite{Tse:Yu:2019,Yu:Wu:2023,Salykina:Khalili:2023,Feng:Zeng:2025}, a dedicated effort is required to develop an appropriate measurement scheme for the effects proposed here.

Our work challenges one of the cornerstone predictions of general relativity, namely the existence of a universal gravitational redshift for photons independent of their state. 
It therefore contributes to the ongoing efforts to understand the interplay of gravity and quantum mechanics, from studies of graviton-mediated interaction of quantum systems \cite{Bertin:Pimentel:2011,Tobar:Manikandan:2024,Hsiang:Cho:2024}, photonics in curved spacetime \cite{Bruschi:Ralph:2014,Exirifard:Karimi:2021,Exirifard:Karimi:2022,Hodgson:Southall:2022,Waite:Hodgson:2025}, and wave-packets dynamics in high-energy physics \cite{Unruh:1995,Casadio:Mersini:2004}, to experiments aimed at testing the quantummn nature of gravity \cite{Amelino-Camelia:1999,Ali:Saurya:2011,Marletto:Vedral:2017,Bose:Mazumdar:2017,Carney2021,Lami:Pedernales:2024,Hanif2024,Liu:Chen:2025,Bose:Fuentes:2025,Plavala:2026}. 

\vspace{0.4cm}

{\footnotesize 

\noindent \textbf{METHODS}

Linearized gravity is employed to describe scenarios where curvature is weak and deviations from flat spacetime are small \cite{Flanagan_2005}. The metric $g_{\mu\nu}$ has the expression $g_{\mu\nu}=\eta_{\mu\nu}+ \varepsilon\, h_{\mu\nu}$, where $\varepsilon\ll1$ is a control parameter and $\eta_{\mu\nu}=\textrm{diag}(-1,1,1,1)$ is the Minkowski metric. Einstein's field equations in vacuum read $\Box\left(h_{\mu\nu}-\frac{1}{2}\eta_{\mu\nu}h\right)=\Box\gamma_{\mu\nu}=0$, which implies that the tensor perturbation $\gamma_{\mu\nu}:= h_{\mu\nu}-\frac{1}{2}\eta_{\mu\nu}h$ propagates as a wave in vacuum. 
Einstein equations contain unphysical degrees of freedom \cite{Misner:1973prb}, which we remove by imposing the harmonic gauge $\partial^\mu\gamma_{\mu\nu}=0$. This allows us to start from the Einstein Hilbert action and derive the Lagrangian density $\mathcal{L}_G=-\frac{1}{4}\left(\partial_\rho\gamma_{\mu\nu}\partial^\rho\gamma^{\mu\nu}-\frac{1}{2}\partial_\rho\gamma\partial^\rho\gamma\right)$, where $\gamma:=\textrm{Tr}(\gamma_{\mu\nu})$, see \cite{Suraj_N_Gupta_1952}.
The variables $\gamma$ and $\gamma_{\mu\nu}$ are treated as independent degrees of freedom for which $\Box\gamma_{\mu\nu}=\Box\gamma=0$.\footnote{An introduction to the methods is found in the literature \cite{Suraj_N_Gupta_1952,Birrell_Davies_1982}. We employ natural units $c=\hbar=1$. $G$ is Newton's constant.}

We now promote $\gamma_{\mu\nu}$ and $\gamma$ to operators. Following standard procedure for canonical quantization we obtain
\begin{subequations}
\begin{align}
\label{spin 2 graviton expansion}
\hat{\gamma}_{\mu\nu}=&\sqrt{8\pi G}\int  \textrm{d}_*^3k\left(\hat{P}_{\mu\nu}(\mathbf{k})e^{i\mathbf{k}\cdot\boldsymbol{x}}+\textrm{h.c.}\right),\\
\label{spin 0 graviton expansion}
    \hat{\gamma}=&\sqrt{16\pi G}\int \textrm{d}_*^3k\left(\hat{P}(\mathbf{k})e^{i\mathbf{k}\cdot\boldsymbol{x}}+\textrm{h.c.}\right)\,,
\end{align}
\end{subequations}
where $\textrm{d}_*^3k:=\textrm{d}^3k/\sqrt{(2\pi)^3 2\omega_{\mathbf{k}}}$, $\omega_{\mathbf{k}}\equiv|\mathbf{k}|$ for the massless  graviton, and $\hat{P}_{\mu\nu}$ and $\hat{P}$ are interpreted as the annihilation operators of free gravitons with spin $2$ and $0$ respectively. They satisfy the commutation relations 

\begin{subequations}
\begin{alignat}{4}\label{graviton algebra 1}
    \bigl[\hat{P}_{\mu\nu}(\mathbf{k}),\hat{P}^\dagger_{\mu'\nu'}(\mathbf{k}')\bigr]=&(\eta_{\mu\mu'}\eta_{\nu\nu'}+\eta_{\mu\nu'}\eta_{\mu'\nu})\delta^3(\mathbf{k}-\mathbf{k}'),\\
\label{graviton algebra 2}
    \bigl[\hat{P}(\mathbf{k}),\hat{P}^\dagger(\mathbf{k}')\bigr]=&-\delta^3(\mathbf{k}-\mathbf{k}')\,\,,
\end{alignat}
\end{subequations}
while all others vanish. The normal-ordered free Hamiltonian of free gravitons is $\hat{H}_{0,\textrm{G}}=\frac{1}{2}\int \textrm{d}^3k|\mathbf{k}|\bigl(\hat{P}_{\mu\nu}^\dagger(\mathbf{k})\hat{P}^{\mu\nu}(\mathbf{k})-2\hat{P}^\dagger(\mathbf{k})\hat{P}(\mathbf{k})\bigr)$.

Within the framework of perturbation theory it is convenient to employ the free field expansions
\begin{subequations}
\begin{align}\label{quantum scalar field}
    \hat{\phi}(x^\mu)=&\int \textrm{d}_*^3k\left[\hat{a}_{\mathbf{k}}e^{i k_\mu x^\mu}+\hat{a}_\mathbf{k}^\dagger e^{-i k_\mu x^\mu}\right].\\
        \hat{\psi}(x^\mu)=&\int \textrm{d}_*^3k\left[\hat{c}_{\boldsymbol{k}}e^{i k_\mu x^\mu}+\hat{c}^\dag_{\boldsymbol{k}}e^{-i k_\mu x^\mu}\right],
    \end{align}
    \end{subequations}
    for the photonic field and the ``source'' field respectively, 
    where $x^\mu\equiv(t,\boldsymbol{x})$, $k^\mu\equiv(\omega_{\boldsymbol{k}},\mathbf{k})$, and $x^\mu k_\mu:=-\omega_{\boldsymbol{k}}t+\boldsymbol{x}\cdot\boldsymbol{k}$. Note that $k^0=|\boldsymbol{k}|$ for the photonic field $\hat{\phi}$, while $k^0=\omega_{\boldsymbol{k}}:=\sqrt{m^2+|{\boldsymbol{k}}|^2}$ for the massive field $\psi$.
Here the sharp momentum operators satisfy the canonical commutation relations $\bigl[\hat{a}_\mathbf{k},\hat{a}_{\mathbf{k}'}^\dagger\bigr]=\bigl[\hat{c}_\mathbf{k},\hat{c}_{\mathbf{k}'}^\dagger\bigr]=\delta^3(\mathbf{k}-\mathbf{k}')$, while all others vanish. The free Hamiltonians for the uncoupled fields $\hat{\phi}$ and $\hat{\psi}$ read $\hat{H}_{0,\phi}=\int \textrm{d}^3k|\mathbf{k}|\hat{a}^\dagger_\mathbf{k}\hat{a}_\mathbf{k}$ and $\hat{H}_{0,\odot}=\int \textrm{d}^3k\omega_{\boldsymbol{k}}\hat{c}^\dagger_\mathbf{k}\hat{c}_\mathbf{k}$ respectively.\footnote{The $\odot$ notation is used for astrophysical objects, such as the sun.}

The interaction between gravitons and other fields is given by
\begin{equation}\label{interaction:Hamiltonian}
    \hat{H}_{\textrm{I},K}=-\frac{1}{2}\int \textrm{d}^3x\left(\hat{\gamma}_{\mu\nu}-\frac{1}{2}\hat{\gamma}\,\eta_{\mu\nu}\right):\hat{T}^{\mu\nu}_K:\,
\end{equation}
where $K=\phi,\odot$ for the photon field $\phi$ and the source field $\psi$ respectively \cite{Suraj_N_Gupta_1952,Bose_2022}. Here $\hat{T}_{\mu\nu}^{\phi}=\partial_\mu\hat{\phi}\partial_\nu\hat{\phi}-\frac{1}{2}\eta_{\mu\nu}\partial^\rho\hat{\phi}\partial_\rho\hat{\phi}$ is the stress-energy tensor for the photons and $\hat{T}_{\mu\nu}^\odot=\partial_\mu\hat{\psi}\partial_\nu\hat{\psi}-\frac{1}{2}\eta_{\mu\nu}(\partial^\rho\hat{\psi}\partial_\rho\hat{\psi}+m^2\hat{\psi}^2)$ is the one for the source. We use the notation $\hat{T}_{\mu\nu}^K$ and $\hat{T}^{\mu\nu}_K$ as needed. The stress-energy tensors are normal ordered to avoid spurious divergences \cite{Srednicki:2007qs}, and are evaluated at $t_0=0$. Note that this perturbative approach requires the interaction between the fields to be small compared to the bare energy scales. This condition is standard in perturbative approaches to quantum field theory \cite{Srednicki:2007qs}, where the backreaction of photons on the background spacetime is negligible, and gravity is weak, i.e., the photon-graviton dynamics occurs far enough from the planet \cite{hayashi1980gravity}. 

In order to model the planet as a classical object,
we consider the extended operators $\hat{C}:=\int \textrm{d}^3k\, F_\odot(\boldsymbol{k}) \hat{c}_{\boldsymbol{k}}$, where $[\hat{C},\hat{C}^\dag]=\int \textrm{d}^3k\, |F_\odot(\boldsymbol{k})|^2=1$. Here $F_\odot$ is the distribution of momenta of the excitations, which is peaked around $\boldsymbol{k}_\odot$ with variance $\sigma_\odot$.
The spatial profile of the source excitations is given by $\tilde{F}_\odot(\boldsymbol{x}):=(2\pi)^{-3}\int \textrm{d}^3 k F_\odot(\boldsymbol{k}) e^{-i \boldsymbol{k}\cdot\boldsymbol{x}}$, which gives us their spatial density $\varrho_\odot(\boldsymbol{x}):=|\tilde{F}_\odot(\boldsymbol{x})|^2$, normalized by $\int \textrm{d}^3x\, \varrho_\odot(\boldsymbol{x})=1$. Since the momentum distribution $F_\odot$ has natural width $\sigma_\odot$, the spatial amplitude $\tilde{F}_\odot$ has natural width $r_\odot:=1/\sigma_\odot$. 
Finally, the source will be assumed to be static, i.e., $\boldsymbol{k}_0=0$, which implies that we will work in the regime $\frac{\lambda_\textrm{c}}{r_\odot}\ll1$, where $\lambda_\textrm{c}=1/m$ is the Compton wavelength of the excitations and $\omega_{\boldsymbol{k}}\approx m$.

The source field is initially found in a coherent state $|\Psi_\odot(0)\rangle:=\hat{D}(\xi)|0\rangle$ defined by $\hat{C}|\Psi_\odot(0)\rangle=\xi|\Psi_\odot(0)\rangle$, where $\hat{D}(\xi):=e^{-(\xi^*\hat{C}-\xi\hat{C}^\dag)}$ is the displacement operator that satisfies $\hat{D}^\dag(\xi)\hat{C}\hat{D}(\xi)=\hat{C}+\xi^*$. 
When $|\xi|\gg1$, the expectation value $T_{\mu\nu}^\odot(\xi):=\langle \Psi_\odot|:\hat{T}_{\mu\nu}^\odot:|\Psi_\odot\rangle$ 
of the stress-energy tensor $:\hat{T}_{\mu\nu}^\odot:$ satisfies the approximation
\begin{align}\label{classical:planet:stress-energy:tensor}
   T_{\mu\nu}^\odot\approx M_\odot\,\varrho_\odot(\boldsymbol{x})\,\delta_{0\mu}\delta_{0\nu},
\end{align}
where we have identified $M_\odot:=|\xi|^2 m$ as the total mass of the planet.
The total energy of the planet is given, to lowest order, by $E_{0,\psi}:=\langle\xi|\hat{H}_{0,\psi}|\xi\rangle\approx M_\odot$, as expected. The stress-energy tensor \eqref{classical:planet:stress-energy:tensor} models a static object with energy density $M_\odot\times\varrho_\odot(\boldsymbol{x})$.

We now compute the evolution of the reduced state $\hat{\rho}_\phi(t):=\textrm{Tr}_{\textrm{G},\odot}\bigl(\hat{U}(t)\hat{\rho}(0)\hat{U}^\dag(t)\bigr)$,
which is obtained by tracing out the degrees of freedom of the gravitons (labelled by G) and the planet (labelled by $\odot$).
The time evolution operator reads $\hat{U}(t):=\exp\bigl[-i \hat{H} t\bigr]$, where $\hat{H}:=\hat{H}_0+\hat{H}_\textrm{I}$ which needs to be supplemented with the following decompositions
\begin{subequations}
\begin{align}
\hat{H}_0=&\hat{H}_{0,\textrm{G}}+\hat{H}_{0,\odot}+\hat{H}_{0,\phi},\\
\hat{H}_\textrm{I} =&\hat{H}_{\textrm{I},\textrm{G}}+\hat{H}_{\textrm{I},\odot},
\end{align}
\end{subequations}
for the free Hamiltonian $\hat{H}_0$ and the interaction Hamiltonian $\hat{H}_\textrm{I}$ respectively.

We introduce the following preliminary notation: the interaction Hamiltonian $\hat{H}_\textrm{I}(t):=\hat{U}_0^\dag(t)\hat{H}_\textrm{I}\hat{U}_0(t)$ in the interaction picture, as well as $\hat{H}_{\textrm{I},K}(t):=\hat{U}_0^\dag(t)\hat{H}_{\textrm{I},K}\hat{U}_0(t)$, where $K=\textrm{G},\odot$; the free evolution operator $\hat{U}_{0,K}(t):=e^{-i \hat{H}_{0,K}t}$ for any subsector, where $K=\textrm{G},\odot,\phi$; the interaction evolution operator $\hat{U}_\textrm{I}(t):=\overset{\leftarrow}{\mathcal{T}}\exp\left[-i\int_0^t\,\textrm{d}t'\,\hat{H}_{\textrm{I}}(t')\right]$.
We employ algebra and in particular the cyclicity of the trace within the space of tracing-out operation to eliminate the $\hat{D}_\odot(\xi\bigr)$ operators, as well as the form $\hat{\rho}_0(0)=|0\rangle_\text{G}\langle0|\otimes|0\rangle_\odot\langle0|\otimes\hat{\rho}_\psi(0)$  of the initial state to find
\begin{align}\label{photon:reduced:state:intermediate:one:appendix}
    \hat{\rho}_\phi(t)
    =&\hat{U}_{0,\phi}(t)\textrm{Tr}_{\textrm{G},\odot}\bigl(\hat{U}_\textrm{I}(\xi,t)\hat{\rho}_0(0)\hat{U}^\dag_\textrm{I}(\xi,t)\bigr)\hat{U}_{0,\phi}^\dag(t)
\end{align}
where we have introduced the operator $\hat{U}_\textrm{I}(\xi,t):=\hat{D}^\dag_\odot(\xi)\hat{U}_\textrm{I}(t)\hat{D}_\odot(\xi)$.
This expression has been obtained using 
\begin{align*}
    \hat{D}^\dag_\odot(\xi)\hat{U}_\textrm{I}(t)\hat{D}_\odot(\xi)
    =\overset{\leftarrow}{\mathcal{T}}\exp\left[-i\int_0^t\,\textrm{d}t'\,\hat{D}^\dag_\odot(\xi)\hat{H}_{\textrm{I}}(t')\hat{D}_\odot(\xi)\right],
\end{align*}
which can occur since $\hat{D}_\odot(\xi)$ is unitary and time-independent. Introducing $ \hat{H}_\textrm{I}(\xi,t):=\hat{D}^\dag_\odot(\xi)\hat{H}_\textrm{I}(t)\hat{D}_\odot(\xi)$, we see that it reads
\begin{align*}
    \hat{H}_\textrm{I}(\xi,t)
    =&\hat{H}_{\textrm{I},\phi}(\xi,t)+\frac{1}{2}\int \textrm{d}^3x\,\hat{h}_{\mu\nu}(t,\boldsymbol{x})\hat{D}^\dag_\odot(\xi):\hat{T}^{\mu\nu}_\odot(t,\boldsymbol{x}):\hat{D}_\odot(\xi).
\end{align*}
We now recall that the (normal-ordered) stress-energy tensor $:\hat{T}^{\mu\nu}_\odot(t,\boldsymbol{x}):$ is quadratic in the annihilation and creation operators $\hat{b}_{\boldsymbol{k}}$ and $\hat{b}_{\boldsymbol{k}}^\dag$, and that $\hat{D}^\dag_\odot(\xi)\hat{b}_{\boldsymbol{k}}\hat{D}_\odot(\xi)=\hat{b}_{\boldsymbol{k}}+\xi F_\odot(\boldsymbol{k})$. Thus we write
\begin{align}
\hat{D}^\dag_\odot(\xi):\hat{T}^{\mu\nu}_\odot(t,\boldsymbol{x}):\hat{D}_\odot(\xi)=&:\hat{T}^{\mu\nu}_\odot(t,\boldsymbol{x}):+|\xi|:\hat{T}^{\mu\nu}_{1,\odot}(t,\boldsymbol{x}):\\\nonumber
&+|\xi|^2T^{\mu\nu}_{2,\odot}(t,\boldsymbol{x}),
\end{align}
where $T^{\mu\nu}_{2,\odot}(t,\boldsymbol{x})$ is obtained from $:\hat{T}^{\mu\nu}_\odot(t,\boldsymbol{x}):$ by replacing each operator $\hat{b}_{\boldsymbol{k}}$ with $\xi F^*_\odot(\boldsymbol{k})$ (and analogously for $\hat{b}_{\boldsymbol{k}}^\dag$), while $:\hat{T}^{\mu\nu}_{1,\odot}(t,\boldsymbol{x}):$ is obtained from $:\hat{T}^{\mu\nu}_\odot(t,\boldsymbol{x}):$ by replacing only one operator 
$\hat{b}_{\boldsymbol{k}}$ or $\hat{b}_{\boldsymbol{k}}^\dag$ in each product with $\frac{\xi}{|\xi|} F^*_\odot(\boldsymbol{k})$ or $\frac{\xi^*}{|\xi|}F_\odot(\boldsymbol{k})$ respectively.
Then
\begin{align*}
    \hat{U}_\textrm{I}(\xi,t)
    =&\hat{U}_{\textrm{I},\odot}(\xi,t) \overset{\leftarrow}{\mathcal{T}}e^{-\frac{i}{2}\int_0^t \textrm{d}t'\int\textrm{d}^3x\, \hat{h}^\odot_{\mu\nu}(t',\boldsymbol{x}):\hat{T}^{\mu\nu}_\phi(t',\boldsymbol{x}):},
\end{align*}
where we have introduced $\hat{h}^\odot_{\mu\nu}(t,\boldsymbol{x}):=\hat{U}_{\textrm{I},\odot}^\dag(\xi,t) \hat{h}_{\mu\nu}(t,\boldsymbol{x})\hat{U}_{\textrm{I},\odot}(\xi,t)$ as well as   
\begin{align}
    \hat{U}_{\textrm{I},\odot}(\xi,t):=\overset{\leftarrow}{\mathcal{T}}\exp\left[-\right.&\frac{i}{2}\int_0^t\,\textrm{d}t'\,\int\textrm{d}^3x\,\hat{h}_{\mu\nu}(t',\boldsymbol{x})\left(|\xi|^2T^{\mu\nu}_{2,\odot}(t',\boldsymbol{x})\right.\nonumber\\
    &\left.\left.+|\xi|:\hat{T}^{\mu\nu}_{1,\odot}(t',\boldsymbol{x}):+:\hat{T}^{\mu\nu}_\odot(t',\boldsymbol{x}):\right)\right].
\end{align}
One of the last steps here is to compute $\hat{U}_{\textrm{I},\odot}^\dag(\xi,t')\,\hat{h}_{\mu\nu}(t',\boldsymbol{x})\hat{U}_{\textrm{I},\odot}(\xi,t')$. This can be done by employing the following trick: first, we introduce a new operator
\begin{align*}
    \hat{U}_{\textrm{I},\odot}(\xi,t):=\overset{\leftarrow}{\mathcal{T}}\exp\left[\right.&-\frac{i}{2}\int_0^t\,\textrm{d}t'\,\int\textrm{d}^3x\,\hat{h}_{\mu\nu}(t',\boldsymbol{x})\left(|\xi|^2T^{\mu\nu}_{2,\odot}(t',\boldsymbol{x})\right.\nonumber\\
    &\left.\left.+|\xi|:\hat{T}^{\mu\nu}_{1,\odot}(t',\boldsymbol{x}):+:\hat{T}^{\mu\nu}_\odot(t',\boldsymbol{x}):\right)r\right].
\end{align*}
We note that the original operator of interest $\hat{U}_{\textrm{I},\odot}(\xi,t)$ can be recovered as $\hat{U}_{\textrm{I},\odot}(\xi,t)=\hat{U}_{\textrm{I},\odot}(r=1,\xi,t)$. Then, we also introduce
\begin{align*}
    \hat{h}^\odot_{\mu\nu}(r,t,\boldsymbol{x}):=\hat{U}_{\textrm{I},\odot}^\dag(r,\xi,t)\,\hat{h}_{\mu\nu}(t,\boldsymbol{x})\hat{U}_{\textrm{I},\odot}(r,\xi,t),
\end{align*}
which allows us to recover the original object simply by computing $\hat{h}^\odot_{\mu\nu}(t,\boldsymbol{x})=\hat{h}^\odot_{\mu\nu}(r=1,t,\boldsymbol{x})$.
We then compute: $\frac{\mathrm{d}}{\mathrm{d}r}\hat{h}_{\mu\nu}(r,t,\boldsymbol{x})$, which can be simplified using the expression $\left[\hat{h}_{\mu\nu}(t,\boldsymbol{x}),\hat{h}_{\mu'\nu'}(t',\boldsymbol{x}')\right]=2 G i\,\frac{\Gamma_{\mu\nu\mu'\nu'}}{|\boldsymbol{x}-\boldsymbol{x}'|}\,\left[\delta((t-t')+|\boldsymbol{x}-\boldsymbol{x}'|)-\delta((t-t')-|\boldsymbol{x}-\boldsymbol{x}'|)\right]$, where we have defined $\Gamma_{\mu\nu\mu'\nu'}:=-\bigl(\eta_{\mu\nu}\eta_{\mu'\nu'}-\eta_{\mu\mu'}\eta_{\nu\nu'}-\eta_{\mu\nu'}\eta_{\mu'\nu}\bigr)$ and which has been obtained by using the identity $ \int \frac{\text{d}^3k}{(2\pi)^3}\frac{\sin\bigl(|\boldsymbol{k}|t+\boldsymbol{x}\cdot\boldsymbol{k}\bigr)}{|\boldsymbol{k}|}=\frac{1}{4\pi|\boldsymbol{x}|}\left(\delta(|\boldsymbol{x}|-t)-\delta(|\boldsymbol{x}|+t)\right)$ and noting that the only Dirac-delta that gives a nonvanishing contribution is $\delta((t-t')-|\boldsymbol{x}-\boldsymbol{y}|)$, since $t\geq t'$. Lengthy algebra gives us
\begin{align*}
    \frac{\mathrm{d}}{\mathrm{d}r}\hat{h}_{\mu\nu}(r,t,\boldsymbol{x})
    =&-G|\xi|^2\int\mathrm{d}^3y\,\frac{\Gamma_{\mu\nu\mu'\nu'}}{|\boldsymbol{x}-\boldsymbol{y}|}\,T^{\mu'\nu'}_{2,\odot}(t_\text{r},\boldsymbol{y})\vartheta(t_\textrm{r}),
\end{align*}
where $t_\text{r}:=t+|\boldsymbol{x}-\boldsymbol{y}|$ is the \textit{retarded time}, we have used the fact that $T^{\mu'\nu'}_{2,\odot}(t_\text{r},\boldsymbol{y})$ is a real function, and we have retained only the leading order terms proportional to $|\xi|\gg1$. 
This differential condition can be immediately solved using the boundary condition $\hat{h}_{\mu\nu}(r=1,t,\boldsymbol{x})=\hat{h}_{\mu\nu}(t,\boldsymbol{x})$, to give us
\begin{align*}
    \hat{h}_{\mu\nu}(r,t,\boldsymbol{x})=&\hat{h}_{\mu\nu}(t,\boldsymbol{x})-G|\xi|^2\,r\,\int\mathrm{d}^3y\,\frac{\Gamma_{\mu\nu\mu'\nu'}}{|\boldsymbol{x}-\boldsymbol{y}|}\,T^{\mu'\nu'}_{2,\odot}(t_\text{r},\boldsymbol{y})\vartheta(t_\textrm{r})\,.
\end{align*}
Thus, we finally have $\hat{h}^\odot_{\mu\nu}(t,\boldsymbol{x})=\hat{h}_{\mu\nu}(r=1,t,\boldsymbol{x})$, which reads
\begin{align}
   \hat{h}^\odot_{\mu\nu}(t,\boldsymbol{x})=\hat{h}_{\mu\nu}(t,\boldsymbol{x})-G|\xi|^2\int\mathrm{d}^3y\frac{\Gamma_{\mu\nu\mu'\nu'}}{|\boldsymbol{x}-\boldsymbol{y}|}T^{\mu'\nu'}_{2,\odot}(t_\text{r},\boldsymbol{y})\vartheta(t_\textrm{r}).
\end{align}
We return to $\hat{U}_\textrm{I}(\xi,t)$, use the expressions computed above, and find
\begin{align*}
    \hat{U}_\textrm{I}(\xi,t)\approx\hat{U}_{\textrm{I,G},\odot,}(\xi,t)\hat{U}_\textrm{I,eff}(t),
\end{align*}
where $\hat{U}_{\textrm{IG}\odot}(\xi,t):=\overset{\leftarrow}{\mathcal{T}}\exp\left[-\frac{i}{2}|\xi|^2\int_0^t\,\textrm{d}t'\,\int\textrm{d}^3x\,\hat{h}_{\mu\nu}(t',\boldsymbol{x})T^{\mu\nu}_{2,\odot}(t',\boldsymbol{x})\right]$ is a unitary operator that acts only on the graviton-source subsectors, and we have introduced
\begin{align}
    \hat{U}_\textrm{I,eff}(t):= \overset{\leftarrow}{\mathcal{T}}\exp\left[G|\xi|^2\frac{i}{2}\int_0^t\,\textrm{d}t'\,\int\frac{\textrm{d}^3x\textrm{d}^3y}{|\boldsymbol{x}-\boldsymbol{y}|}\,\hat{\Delta}(t',t_\text{r},\boldsymbol{x},\boldsymbol{y})\right],
\end{align}
where $\hat{\Delta}(t',t_\text{r},\boldsymbol{x},\boldsymbol{y}):=2T^{\mu\nu}_{2,\odot}(t_\text{r},\boldsymbol{y}):\hat{T}_{\mu\nu}^\phi(t',\boldsymbol{x}):-T_{2,\odot}(t_\text{r},\boldsymbol{y}):\hat{T}_\phi(t',\boldsymbol{x}):\vartheta(t_\textrm{r})$ for simplicity of presentation.

The object $T^{\mu\nu}_{2,\odot}(t_\text{r},\boldsymbol{y})$ is determined by products of functions $F_\odot(\boldsymbol{k})$. We now observe that the distribution $F_\odot(\boldsymbol{k})$ is localized around the origin $\boldsymbol{k}=0$ with variance $\sigma_\odot$. In order to have a source of gravitons that remains static and localized, we also need to assume that $\sigma_\odot$ is sufficiently small, i.e. $\sigma_\odot\lambda_\text{c}\ll1$ where $\lambda_\text{c}=1/m$ is the Compton wavelength of the excitations. In this regime, one has $\omega_{\boldsymbol{k}}=\sqrt{m^2+|\boldsymbol{k}|}\approx m(1+\mathcal{O}((\sigma_\odot\lambda_\text{c})^2))$. It is easy to see that the components $T_{0j}^\odot(\xi)$ and $T_{jk}^\odot(\xi)$ will be much smaller than $T_{00}^\odot(\xi)$, since the former are defined by integrals that contain linear or quadratic terms in the momentum, while the latter is quadratic in the mass $m$. Thus, we have $\frac{1}{m^2}T_{00}^\odot(\xi)\approx\mathcal{O}(1)$, $T_{0j}^\odot(\xi)\approx\mathcal{O}(\sigma_\odot/m)$ and $T_{jk}^\odot(\xi)\approx\mathcal{O}((\sigma_\odot/m)^2)$, and conclude that
\begin{align*}
    T^{\mu\nu}_\odot(\xi)\approx\frac{1}{2}((\partial_0\hat{\psi})^2+m^2\hat{\psi}^2)|_{\hat{b}=b}\delta^\mu{}_0\delta^\nu{}_0
\end{align*}
to lowest order in $\sigma/m$. The only caveat that needs to be considered is the fact that we have approximated $e^{-i\omega_{\boldsymbol{k}}t}\approx e^{-mit}$, which is valid up to $\mathcal{O}((\sigma_\odot\lambda_\text{c})^2)$ for times $t\ll \hbar m/p_\odot^2$, where $p_\odot:=\hbar\sigma_\odot$ is the variance of the momentum distribution. Here we have restored units for the sake of estimating this bound. Given a mass $m=10^{-23}$kg, and a momentum variance $p_\odot=10^{-30}$kg m/s, one obtains that the approximation is valid as long as $t\ll10^{7}$s$\approx3\times0.3$y. In this regime we also have that $\sigma_\odot\lambda_\text{c}=(\sigma_\odot\hbar)/(mc)\approx10^{-15}\ll1$ as required by the approximation. These numbers give only a crude estimate, since a realistic planet is held together by forces that are not considered here. More work would be necessary to estimate correctly in which approximation is our scenario based.

We now need to replace each operator $\hat{b}_{\boldsymbol{k}}$ with $\xi F^*(\boldsymbol{k})$ as concluded above and use the fact that $\omega_{\boldsymbol{k}}\approx m$, to obtain
\begin{subequations}
    \begin{align}
    \partial_0\hat{\psi}\to&
    =-i(2\pi)^{3/2}\sqrt{\frac{m}{2}}(\xi\,e^{-i m t}-\xi^* e^{i m t})\tilde{F}_\odot(\boldsymbol{x}),\\
    \hat{\psi}\to&
    =\frac{(2\pi)^{3/2}}{\sqrt{2m}}(\xi\,e^{-i m t}+\xi^* e^{i m t})\tilde{F}_\odot(\boldsymbol{x}).
\end{align}
\end{subequations}
We can now combine these expressions with that for $\hat{T}^{\mu\nu}_\odot(\xi)$ in terms of the field obtained before, to finally have
\begin{align}
   T_{\mu\nu}^\odot\approx M_\odot\,\varrho_\odot(\boldsymbol{x})\,\delta_{0\mu}\delta_{0\nu},
\end{align}
where we have identified
\begin{align}
    M_\odot:=|\xi|^2 m
\end{align}
as the total mass of the massive object and we recall that $\varrho_\odot(\boldsymbol{x}):=(2\pi)^{3}|\tilde{F}_\odot(\boldsymbol{x})|^2$ is the classical probability density of the source field normalized by $\int \text{d}^3x\varrho_\odot(\boldsymbol{x})=1$.

Using the fact that $2\hat{T}^\phi_{00}+2\hat{T}^\phi=2(\partial_0\hat{\phi})^2$, we find
\begin{align}\label{effective:interaction:unitary:operator:appendix}
    \hat{U}_\textrm{I,eff}(t):= \overset{\leftarrow}{\mathcal{T}}e^{\frac{r_\textrm{S}}{2} i\int_0^t\,\textrm{d}t'\,\int\frac{\textrm{d}^3x\textrm{d}^3y}{|\boldsymbol{x}-\boldsymbol{y}|}\, \varrho_\odot(\boldsymbol{y})\left(:\partial_0\phi(t',\boldsymbol{x}):\right)^2\vartheta(t_\textrm{r})},
\end{align}
where $r_\textrm{S}:=2GM_\odot$ is the Schwarzschild radius of the planet.

These result give us $\hat{\rho}_\phi(t)
    \approx\hat{U}_{0,\phi}(t)\hat{U}_\textrm{I,eff}(t)\hat{\rho}_\phi(0)\hat{U}_\textrm{I,eff}^\dag(t)\hat{U}_{0,\phi}^\dag(t)$, which in turn allows us to write
\begin{align}\label{photon:reduced:state:final:appendix}
    \hat{\rho}_\phi(t)
    \approx&\hat{U}_\textrm{eff}(t)\hat{\rho}_\phi(0)\hat{U}_\textrm{eff}^\dag(t),
\end{align}
where we have defined
\begin{subequations}\label{final:evolution:operator:appendix}
\begin{align}
    \hat{U}_\textrm{eff}(t):=&\overset{\leftarrow}{\mathcal{T}}\exp\left[ -i\int_0^t\,\textrm{d}t'\,\left(\hat{H}_{0,\phi}+\hat{H}_{2,\phi}(t')\right)\right],\\
    \hat{H}_{2,\phi}(t):=&-\frac{r_\textrm{S}}{2}\int\frac{\textrm{d}^3x\textrm{d}^3y}{|\boldsymbol{x}-\boldsymbol{y}|}\, \varrho_\odot(\boldsymbol{y})\left(:\partial_0\phi(0,\boldsymbol{x}):\right)^2\vartheta(t_\textrm{r}),
\end{align}
\end{subequations}
as well as $\partial_0\phi(0,\boldsymbol{x})=\partial_0\phi(t=0,\boldsymbol{x})$.
We conclude this part by noting that \eqref{photon:reduced:state:final:appendix} and \eqref{final:evolution:operator:appendix} are the key expressions for this work.

We specialize the main results \eqref{photon:reduced:state:final:appendix} and \eqref{final:evolution:operator:appendix} to the case of a spherical planet. 
Performing the integral over $\boldsymbol{x}$ in polar coordinates of the Hamiltonian in \eqref{final:evolution:operator:appendix} gives us
\begin{align}\label{effective:photonic:hamiltonian:appendix}
\hat{H}_{2,\phi}(t)=&\frac{r_\textrm{S}}{4}\int \frac{\textrm{d}^3k\textrm{d}^3k'}{(2\pi)^3}K_{\boldsymbol{k}\boldsymbol{k}'}(t)\bigl(2\hat{a}^\dagger_{\boldsymbol{k}}\hat{a}_{\boldsymbol{k}'}-\hat{a}_{\boldsymbol{k}}\hat{a}_{-\boldsymbol{k}'}-\textrm{h.c.}\bigr),
\end{align}
where we have introduced the function
\begin{align}\label{K:function:appendix}
    K_{\boldsymbol{k}\boldsymbol{k}'}(t):=-\frac{\sqrt{|\boldsymbol{k}||\boldsymbol{k}'|}}{\pi^2}\frac{\sin^2\bigl(|\boldsymbol{k}-\boldsymbol{k}'|t/2\bigr)}{|\boldsymbol{k}-\boldsymbol{k}'|^2}\tilde{\rho}_\odot(\boldsymbol{k}-\boldsymbol{k}').
\end{align}
as well as he momentum distribution of the source excitations $\tilde{\rho}_\odot(\boldsymbol{k})=\int \textrm{d}^3x\varrho_\odot(\boldsymbol{x})e^{i\boldsymbol{x}\cdot\boldsymbol{k}}$ (where we assume that $\tilde{\rho}_\odot(\boldsymbol{k})=\tilde{\rho}_\odot^*(\boldsymbol{k})$).

When the dynamics of interest occur far away from the planet, that it can be approximated as a pointlike object. In this case one has $\varrho_\odot(\boldsymbol{x})=\delta^3(\boldsymbol{x})$, and therefore we also have $\tilde{\rho}_\odot(\boldsymbol{k})=1$. We assume to be working in this regime here.

The Hamiltonian \eqref{effective:photonic:hamiltonian:appendix} is quadratic in the annihilation and creation operators $\hat{a}_{\boldsymbol{k}},\hat{a}^\dag_{\boldsymbol{k}}$. Therefore, it induces \emph{linear} dynamics in the sense of dynamics that preserve the Gaussian character of the state \cite{Adesso:Ragy:2014}. This is also true for the full effective Hamiltonian $\hat{H}_\textrm{eff}:=\hat{H}_{0,\phi}+\hat{H}_{2,\phi}$, and we can therefore write
\begin{align}
\hat{U}_\textrm{eff}^\dag(t)\hat{a}_{\boldsymbol{k}}\hat{U}_\textrm{eff}(t)=\int \textrm{d}^3k' \left(\alpha_{\boldsymbol{k}\boldsymbol{k}'}(t)\hat{a}_{\boldsymbol{k}'}+\beta_{\boldsymbol{k}\boldsymbol{k}'}(t)\hat{a}^\dag_{\boldsymbol{k}'}\right),
\end{align}
which is known as a \emph{Bogoliubov transformation}, where $\alpha_{\boldsymbol{k}\boldsymbol{k}'}(t),\beta_{\boldsymbol{k}\boldsymbol{k}'}(t)$ are the Bogoliubov coefficients \cite{Birrell_Davies_1982}. These coefficients satisfy the Bogoliubov identities $(\boldsymbol{\alpha}\boldsymbol{\alpha}^\dag)_{\boldsymbol{k}\boldsymbol{k}'}-(\boldsymbol{\beta}\boldsymbol{\beta}^\dag)_{\boldsymbol{k}\boldsymbol{k}'}=\delta^3(\boldsymbol{k}-\boldsymbol{k}')$ and $\boldsymbol{\alpha}^*\boldsymbol{\alpha}^\textrm{Tp}-\boldsymbol{\beta}^*\boldsymbol{\beta}^\textrm{Tp}=0$ in matrix form \footnote{We ignore here the problem of formally defining these transformations in a space that has an infinite ans continuous basis.}.

The Gaussian channel determined above is plagued by the following issue: the overlap $p_0(t):=|\langle0|\hat{U}_\textrm{eff}(t)|0\rangle|^2$ is formally divergent by computing $p_0(t)$ to second order in $r_\textrm{S}$. We have
\begin{align*}
    p_0(t)=&1-r_\textrm{S}^2\,t^2 \int \frac{\textrm{d}^3k\textrm{d}^3k'}{8\pi^4(2\pi)^6}\frac{|\boldsymbol{k}-\boldsymbol{k}'||\boldsymbol{k}'|}{|\boldsymbol{k}|^4},
\end{align*}
which in general diverges due to the asymptotic behavior  $|\boldsymbol{k}'|^{4}$ of the integrand for large $|\boldsymbol{k}'|$.

This issue is closely related to the problem of \emph{renormalization}, which we treat as follows. We note that the Hamiltonian \eqref{effective:photonic:hamiltonian:appendix} contains two types of terms: the \emph{mode-mixing} terms $\hat{a}^\dagger_{\boldsymbol{k}}\hat{a}_{\boldsymbol{k}'}\pm \hat{a}^\dagger_{\boldsymbol{k}'}\hat{a}_{\boldsymbol{k}}$, which induce \emph{passive operations} that preserve the total number of excitations, and the \emph{squeezing} terms $\hat{a}_{\boldsymbol{k}}^\dag\hat{a}_{-\boldsymbol{k}'}^\dag\pm\hat{a}_{\boldsymbol{k}}\hat{a}_{-\boldsymbol{k}'}$, which chnage the total number of excitations \cite{Adesso:Ragy:2014}. We can therefore move to the interaction picture and apply the rotating wave approximation to discard the co-rotating squeezing terms \cite{Heib:Lageyre:2025}, which leaves us with the final Hamtilonian
\begin{align}
\hat{H}_{2,\phi}(t)=&r_\textrm{S}\int \frac{\textrm{d}^3k\textrm{d}^3k'}{2\pi^2(2\pi)^3}\mathcal{H}_{\boldsymbol{k},\boldsymbol{k}'}(t)\hat{a}^\dagger_{\boldsymbol{k}}\hat{a}_{\boldsymbol{k}'},
\end{align}
where we have introduced $\mathcal{H}_{\boldsymbol{k},\boldsymbol{k}'}(t)=-\frac{\sqrt{|\boldsymbol{k}||\boldsymbol{k}'|}}{|\boldsymbol{k}-\boldsymbol{k}'|^2}\bigl(1-\cos\bigl(|\boldsymbol{k}-\boldsymbol{k}'|t\bigr)\bigr)$ for convenience.
The full channel given by $\hat{U}_\textrm{eff}(t)$ reduces to
\begin{align}
    \hat{a}_{\boldsymbol{k}}(t)=&\hat{U}^\dag_\textrm{eff}(t)\hat{a}_{\boldsymbol{k}}\hat{U}_\textrm{eff}(t)\int\textrm{d}^3k'\alpha_{\boldsymbol{k}\boldsymbol{k}'}\hat{a}_{\boldsymbol{k}'},
\end{align}
where $\alpha_{\boldsymbol{k}\boldsymbol{k}'}$ are the Bogoliubov coefficients that mix modes without particle creation \cite{Birrell_Davies_1982}. In particular, they must satisfy the (Bogoliubov) constraint $\int\textrm{d}^3k'\alpha_{\boldsymbol{k}\boldsymbol{k}'}\alpha_{\boldsymbol{k}''\boldsymbol{k}'}^*=\delta^3(\boldsymbol{k}-\boldsymbol{k}'')$, which implies that $\boldsymbol{\alpha}$ is formally a unitary matrix, as expected.

All remaining computations, that is, the computations of the main results, can be obtained using this effective time evolution operator and lengthy tedious algebra.
}

\vspace{0.2cm}

\noindent \textbf{ACKNOWLEDGMENTS}

We acknowledge Stefano Mancini and Roberto Casadio for their suggestions and comments. AL is particularly thankful to Salvatore Capozziello for useful discussions to improve the quality of this manuscript, as well as for supporting a long-term visit to Forschungzentrum J\"ulich. 
AF and DEB acknowledge support from the joint project No. 13N15685 ``German Quantum Computer based on Superconducting Qubits (GeQCoS)'' sponsored by the German Federal Ministry of Education and Research (BMBF) under the \href{https://www.quantentechnologien.de/fileadmin/public/Redaktion/Dokumente/PDF/Publikationen/Federal-Government-Framework-Programme-Quantum-technologies-2018-bf-C1.pdf}{framework programme
``Quantum technologies -- from basic research to the market''}. D.E.B. also acknowledges support from the German Federal Ministry of Education and Research via the \href{https://www.quantentechnologien.de/fileadmin/public/Redaktion/Dokumente/PDF/Publikationen/Federal-Government-Framework-Programme-Quantum-technologies-2018-bf-C1.pdf}{framework programme
``Quantum technologies -- from basic research to the market''} under contract number 13N16210 ``SPINNING''.

\vspace{0.2cm}

\noindent \textbf{AUTHOR CONTRIBUTIONS}

DEB conceived the idea. DEB provided the first approach to solving the problem. AL, AF, and DEB revised the approach and improved the results. DEB developed the final approach and revised the techniques used, obtaining the current results. AWS provided crucial insights in how to pose the relevant questions. All authors contributed to writing the manuscript.

\vspace{0.2cm}

\noindent \textbf{NOTES}

After finalizing a major revision of this work, the authors became aware of the independent work \cite{Aziz:Howl:2025}.

\bibliographystyle{unsrt}
\bibliography{main}

\end{document}